IAC-15.B5.1.9

# ON THE INFLUENCE OF IMPACT EFFECT MODELLING FOR GLOBAL ASTEROID IMPACT RISK DISTRIBUTION

Clemens Rumpf [a] (c.rumpf@soton.ac.uk) (corresponding),
Hugh G. Lewis [a] (h.g.lewis@soton.ac.uk),
Peter M. Atkinson [b,c,d,e] (pma@lancaster.ac.uk)

[a] University of Southampton, Faculty of Engineering and Environment, Southampton, UK,
[b] Lancaster University, Faculty of Science and Technology, Lancaster, UK,
[c] University of Utrecht, Faculty of Geosciences, Utrecht, The Netherlands
[d] University of Southampton, Geography and Environment, Southampton, UK
[e] Queen's University Belfast, School of Geography, Archaeology and Palaeoecology, Belfast, UK

## ABSTRACT

The collision of an asteroid with Earth can potentially have significant consequences for the human population. The European and United States space agencies (ESA and NASA) maintain asteroid hazard lists that contain all known asteroids with a non-zero chance of colliding with the Earth in the future. Some software tools exist that are, either, capable of calculating the impact points of those asteroids, or that can estimate the impact effects of a given impact incident. However, no single tool is available that combines both aspects and enables a comprehensive risk analysis. The question is, thus, whether tools that can calculate impact *location* may be used to obtain a qualitative understanding of the asteroid impact *risk* distribution. To answer this question, two impact risk distributions that control for impact effect modelling were generated and compared. The Asteroid Risk Mitigation Optimization and Research (ARMOR) tool, in conjunction with the freely available software OrbFit, was used to project the impact probabilities of listed asteroids with a minimum diameter of 30 m onto the surface of the Earth representing a random sample (15% of all objects) of the hazard list. The resulting 261 impact corridors were visualized on a global map. Furthermore, the impact corridors were combined with Earth population data to estimate the "simplified" risk (without impact effects) and "advanced" risk (with impact effects) associated with the direct asteroid impacts that each nation faces from present to 2100 based on this sample. The relationship between risk and population size was examined for the 40 most populous countries and it was apparent that population size is a good proxy for relative risk. The advanced and simplified risk distributions were compared and the alteration of the results based on the introduction of physical impact effects was discussed. Population remained a valid proxy for relative impact risk, but the inclusion of impact effects resulted in significantly different risks, especially when considered at the national level. Therefore, consideration of physical impact effects is essential in estimating the risk to specific nations of the asteroid threat.

## 1. INTRODUCTION

Earth has collided with asteroids since it was an embryo planet and this process continues today, albeit at a lower rate [1]. Asteroid impacts have been responsible for at least two major disruptions in the evolution of life [2,3] and today, they remain a potential hazard for the human population [4,5]. Surveys scan the sky for asteroids in an effort to discover as many as possible and to calculate their orbits [6]. Based on the propagation of orbits, those asteroids that may potentially impact the Earth in the future are identified. The European Space Agency (ESA) and the National Aeronautics and Space Administration (NASA), perform such collision detection using automated systems and the results are published on their respective Near Earth Object (NEO) webpages [7,8]. The NEO webpages provide information about the orbit, physical properties and impact probability of the listed objects. However, a risk assessment, involving the asteroid's possible impact locations, is not part of the available information. Only in rare cases, such as for asteroid Apophis [9], are impact locations (in the shape of an impact corridor) calculated. Impact locations can help to develop an intuitive understanding of the specific impact risk as it shows the areas that would be affected, but, to gain a reliable understanding for risk, impact effects and their implications for population have to be calculated at the impact locations. The calculation of impact effects has been treated in the literature, and two software tools exist that are capable of estimating impact [10,11]. Although these tools calculate impact effects, they do not routinely close the loop between orbital characteristics that determine impact conditions, probability and location, and, thus, only offer limited utility in understanding impact risk. A comprehensive risk assessment, taking into account the physical impact effects that are governed by the asteroid's speed, impact angle and physical properties on one hand and the distribution of population along the impact corridor on the other hand, is, thus far, not available. Nevertheless, such a risk assessment is needed to help determine the political and technological response when an asteroid with a significant impact probability is discovered [12]. Could existing tools be used to perform such impact risk assessment?

Given that the practice of impact location calculation is already available to some entities within the planetary defense community [13,14], it is of interest to know how well this knowledge of impact locations enables impact risk estimation. The validity of an impact risk estimate that is based on impact location knowledge alone can be gauged through comparison with a similar risk estimate that takes impact effect modelling into account. Following this logic, two risk distributions are needed: A "simplified" one that considers impact location without impact effects and an "advanced" one that accounts for impact location as well as impact effects. To generate a global distribution of asteroid impacts, a sufficiently large set of orbital and physical data of, ideally, real asteroids should be used. Using real asteroid data not only allows calculation of impact locations but also facilitates determination of representative impact conditions such as impact energy and angle. Some variable that is exposed to the asteroid impact hazard needs to be identified to complete the risk estimation process, and the global, human population map represents a suitable dataset that is also publicly available [15].

To meet the objectives mentioned above, the Asteroid Risk Mitigation Optimization and Research (ARMOR) tool was developed. ARMOR calculates the impact locations of asteroids and presents this information in the form of spatial impact probability distributions on the Earth map. Furthermore, the physical conditions of the impact, such as speed and impact angle are determined and facilitate calculation of physical impact effects and interaction with the population on the ground. Taking into account the global impact probability, this information enables risk calculation of asteroids.

## 2. METHOD

This section describes the method of calculating impact locations and their associated impact probabilities, the resulting impact effects and how these components are used to estimate risk.

It should be noted that the term "impact" can refer to a general collision between an asteroid and the Earth, an airburst or the event of contact between the asteroid and the surface of the Earth. Where the context requires clarification, the first case is referred to as "collision" the second as "airburst" and the third case as "ground impact". The method starts with the calculation of the impact location of observed asteroids.

### 2.1 IMPACT LOCATION AND PROBABILITY

The nominal orbital solution of an asteroid is a state vector describing the asteroid's orbit and position that fit best the observations that are available for this asteroid. A covariance matrix represents the uncertainty region that is associated with the orbital solution. The uncertainty region has a weak direction, commonly referred to as the Line of Variation (LOV), along which the asteroid position is only poorly constrained and it typically stretches along the orbit of the asteroid [16]. Using the available observations and the current nominal orbital solution of an asteroid that are provided on the ESA NEO webpage, the freely available software OrbFit [17] was utilized to identify orbit solutions that lie on the LOV as well as inside the uncertainty region and that result in a future Earth impact. OrbFit samples the uncertainty region to find such impacting orbit solutions, which are called virtual impactors (VI).

The ARMOR tool was used to project the impact probability of each VI onto the surface of the Earth. ARMOR used the VI orbit solution from OrbFit as the initial condition for a ten day trajectory propagation that results in an impact. By sampling the LOV close to the original VI orbit solution, many impact points are calculated that form the centre line of the impact probability corridor on the Earth. Impact probability is assumed to be constant along the centre line (valid for low impact probabilities, such as used here), while a normal distribution with a 1-sigma value equal to the LOV width (a parameter available on the NEO webpages) represents the cross track impact probability. The impact probability distribution is scaled such that its integral is equal to the impact probability of the VI. The method of ARMOR, along with validation cases, is presented in greater detail in [18,19]. For practical purposes the term VI will describe the original impacting orbit solution [20] along with its immediate neighbouring uncertainty space that forms the impact corridor on the Earth. It should be noted that one asteroid may have multiple, dynamically separate impact solutions in the future and, thus, yield more than one VI.

The global impact distribution of 69 known asteroids resulting in 261 potential collisions with the Earth was calculated and the result is the basis for the simplified and advanced forms of risk calculation.

## 2.2 IMPACT EFFECTS

Six physical impact effects are associated with an asteroid impact and the occurrence of each effect depends on whether the asteroid frees most of its energy during atmospheric passage or upon ground impact. While passing through the atmosphere, aerodynamic friction heats the asteroid which induces surface ablation [21]. In this environment, smaller asteroids are prone to undergo rapid disintegration in an explosion-like event called an airburst [22] producing an overpressure shock wave, which prompts strong, gust-like winds, that propagate away from the airburst location. The airburst also emits thermal radiation that can burn surfaces which are impinged. Larger asteroids can pass through the atmosphere intact and produce a crater upon land impact [23]. The cratering process itself, as well as the accompanying out-throw of ejecta, account for two additional impact effects, while the ground impact-induced seismic shaking represents the last. Similarly to an airburst, the cratering event produces overpressure, wind gust and thermal radiation. Mathematical models of these impact effects are presented in [10] and ARMOR's impact effect modelling is derived directly from that source. Table 1 shows how the ARMOR implementation of the models for the impact effects replicates the results of the web based "Earth Impact Effects Program" (EIEP) [10] for an arbitrarily selected airburst and a cratering impact. Errors between the two programmes are small suggesting correct implementation of the models and the errors can be attributed to the fact that EIEP values are only available in rounded format, to varying definitions of universal constants, and to incomplete descriptions about EIEP algorithm implementation.

*Table 1: Sample asteroid impact effect calculation for an airburst and ground impact case are presented and results of ARMOR are compared to "Earth Impact Effects Program" (EIEP) [10] output. Note that EIEP values are generally rounded. The results suggest correct implementation of the EIEP routines in ARMOR.*

| Airburst | | |
|---|---|---|
| Distance to Impact | 2 | km |
| Asteroid Diameter | 50 | m |
| Asteroid Density | 3100 | kg/m^3 |
| Asteroid Speed | 22 | km/s |
| Impact Angle | 55 | degree |

| Aspect | ARMOR | EIEP | Error [%] |
|---|---|---|---|
| Kinetic Energy [J] | 4.91E+16 | 4.91E+16 | 0.0 |
| Breakup Altitude [m] | 57128 | 57100 | 0.0 |
| Airburst Altitude [m] | 5115 | 5120 | 0.1 |
| Post Airburst Speed [km/s] | 6.527 | 6.53 | 0.0 |
| Airburst Energy [J] | 4.48E+16 | 4.48E+16 | 0.1 |
| Overpressure [Pa] | 171811 | 172000 | 0.1 |
| Wind Speed [m/s] | 266 | 258 | 3.1 |

| Cratering Impact | | |
|---|---|---|
| Distance to Impact | 10 | km |
| Asteroid Diameter | 250 | m |
| Asteroid Density | 3100 | kg/m^3 |
| Asteroid Speed | 27 | km/s |
| Impact Angle | 35 | degree |
| Ground Density | 2500 | kg/m^3 |

| Aspect | ARMOR | EIEP | Error [%] |
|---|---|---|---|
| Kinetic Energy [J] | 9.24E+18 | 9.24E+18 | 0.0 |
| Breakup Altitude [m] | 60406 | 60400 | 0.0 |
| Impact Speed [km/s] | 23.73 | 23.70 | 0.1 |
| Impact Energy [J] | 7.14E+18 | 7.14E+18 | 0.1 |
| Transient Crater Diameter [km] | 3.92 | 3.92 | 0.1 |
| Final Crater Diameter [km] | 4.90 | 4.71 | 3.9 |
| Thermal Exposure [J/m^2] | 3.32E+07 | 3.40E+07 | 2.3 |
| Seismic Shaking [Richter Mag] | 6.76 | 6.80 | 0.6 |
| Ejecta Thickness [m] | 2.10 | 2.10 | 0.1 |
| Overpressure [Pa] | 1046182 | 1050000 | 0.4 |
| Wind Speed [m/s] | 807 | 781 | 3.4 |

It should be noted that tsunamis were not taken into account in this analysis to maintain comparability between result sets as described in section 2.3. Tsunamis potentially have the farthest reach of all impact effects [6], and, together with the general tendency of populations to concentrate in coastal regions, may constitute a significant contributor to risk calculation. The importance of tsunami modelling is expected to increase with larger sized asteroids [24]. However, simulations showed that because of their sizes (one is 341 m and the rest is sub-200 m), only one third (25 out of 69) of the asteroids produced ground impacts, and, thus, potential tsunamis, in this analysis.

## 2.3 RISK

The risk associated with a map cell ($R_c$) is defined here as the product of the asteroid impact probability in a specific cell ($\rho_c$), the number of people who live in the area who are exposed to the impact-generated effects ($\Psi$) and the vulnerability of the exposed population ($\eta$) to the impact effects that are attenuated by distance from the impact site.

$$R_c = \rho_c \times \Psi \times \eta \qquad Eq\ 1$$

Vulnerability of the population is conditional upon the severity of the impact effects and severity describes how powerful each effect is at a given distance. Very strong impact effects (e.g. magnitude 8-equivalent seismic shaking) are likely to cause more casualties and, thus, are associated with high population vulnerability, while a moderate tremor (e.g. magnitude 4 seismic shaking) is likely to leave the population mostly unharmed and results in low vulnerability. In this way, the hazard magnitude is implicitly accounted for in the above equation.

In the simplified impact risk analysis, which did not account for physical impact effects, vulnerability ($\eta$) was set equal to one because the simulation did not calculate the severity and extent of impact effects. Instead, it was assumed that the impact cell's population is equal to the number of casualties (in other words, if an impact occurs in a cell then everyone in it dies). Risk, in this case was the product of impact cell population times probability of impact in that cell. In contrast, the advanced results account for physical impact effects and, consequently, vulnerability is modelled as a function of severity. Furthermore, the attenuation of impact effects with distance is modelled which allows account of the population in affected cells beyond the impact cell and varying vulnerability as impact effects weaken while they propagate into more distant cells.

Finally, the national risk of all countries is estimated by summing the risk distribution within each country's borders. The national risk values are divided by the global risk to produce the percentage of global risk that each country faces and this is called the relative risk of a nation. Eq 2 is the formula for relative risk in each map cell as a function of total risk in each map cell caused by a specific VI ($R_{i,c}$). Subscript $c$ denotes one cell in the world map, and subscript $i$ identifies the VI.

$$\text{relative risk}_c = \sum_i^{261} \frac{R_{i,c}}{\sum_c R_{i,c}} \qquad Eq\ 2$$

In this way, both, the simplified as well as the advanced risk distribution are normalized with respect to global risk producing relative risk estimates. The estimation of relative risk provides a basis for comparison between the simplified and advanced results. Furthermore, given that the simplified risk estimate can, by definition, account for land impacts only, only direct land impacts (and near coastal impacts in the case of the advanced result) are considered in the analysis to retain comparability between the two risk distributions. This means that asteroid impact-generated tsunamis are not part of the analysis. Relative risk, in contrast to absolute risk, does not retain information on the number of expected casualties through the process of normalization. Consequently, a statement about the magnitude of the expected risk is impossible with this approach. However, relative risk illustrates the risk level of one country in comparison with another and allows statements about which country is more at risk based on the given impact probability distribution.

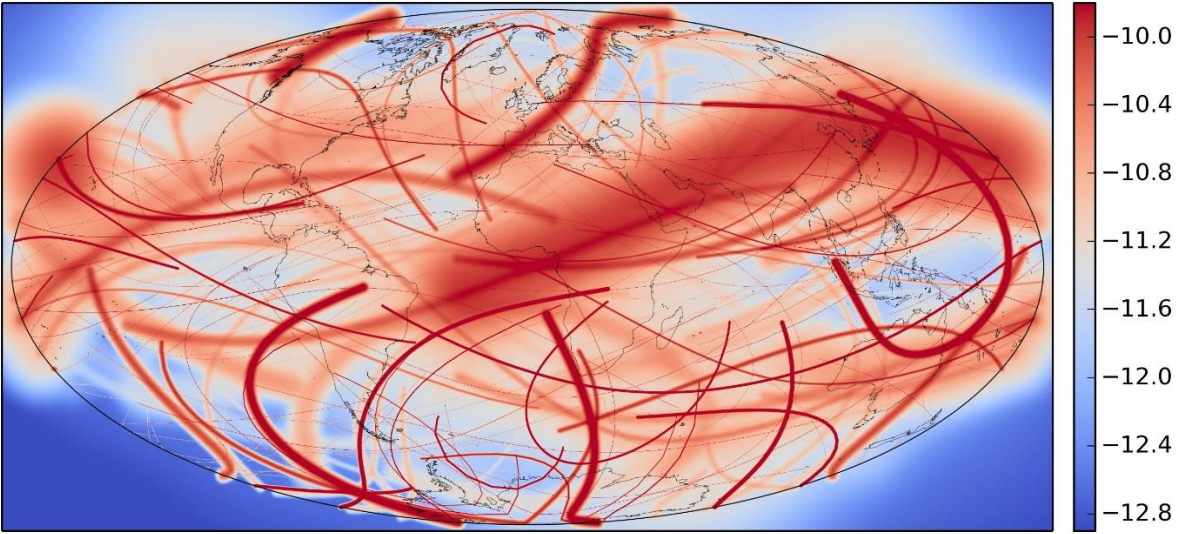

*Figure 1: The Earth in the Hammer projection showing the impact probability distributions for 261 VIs. The colour coding represents the impact probability at each location using a logarithmic scale (shown right).*

## 3. RESULTS

The following results are based on 69 known asteroids that produce 261 VIs, and these asteroids were sampled in the November 2014 timeframe. All 261 VIs were subjected to the method described above to produce a set of impact corridors, each in the form of a Gaussian distribution that reflects the impact probabilities of the assessed VIs. All impact solutions were combined within a global map and the global impact probability distribution is shown in Figure 1. Based on the combined probability corridors in this map, the impact probability in each map cell was determined for subsequent risk calculation.

By means of combining Figure 1 with the world population, the global risk distribution without consideration for physical impact effects was produced and this is the simplified risk distribution (Figure 2) [25].

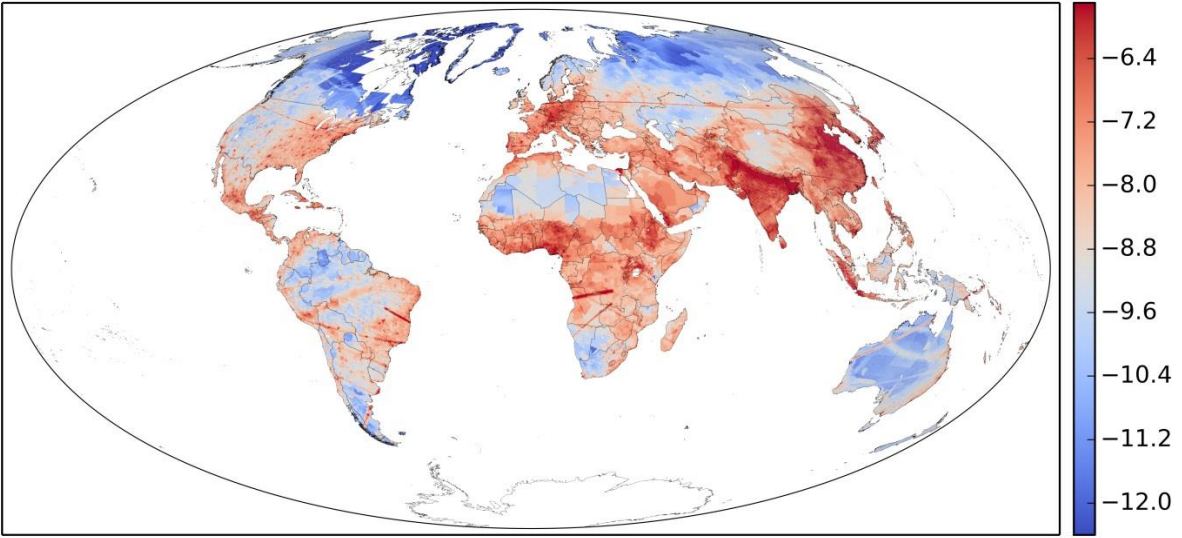

*Figure 2: Simplified asteroid risk map. This is the product of impact probability and world population but does not account for physical impact effects. The colour in each region indicates the risk level. Risk is normalized with respect to global risk and is colour coded using a logarithmic scale.*

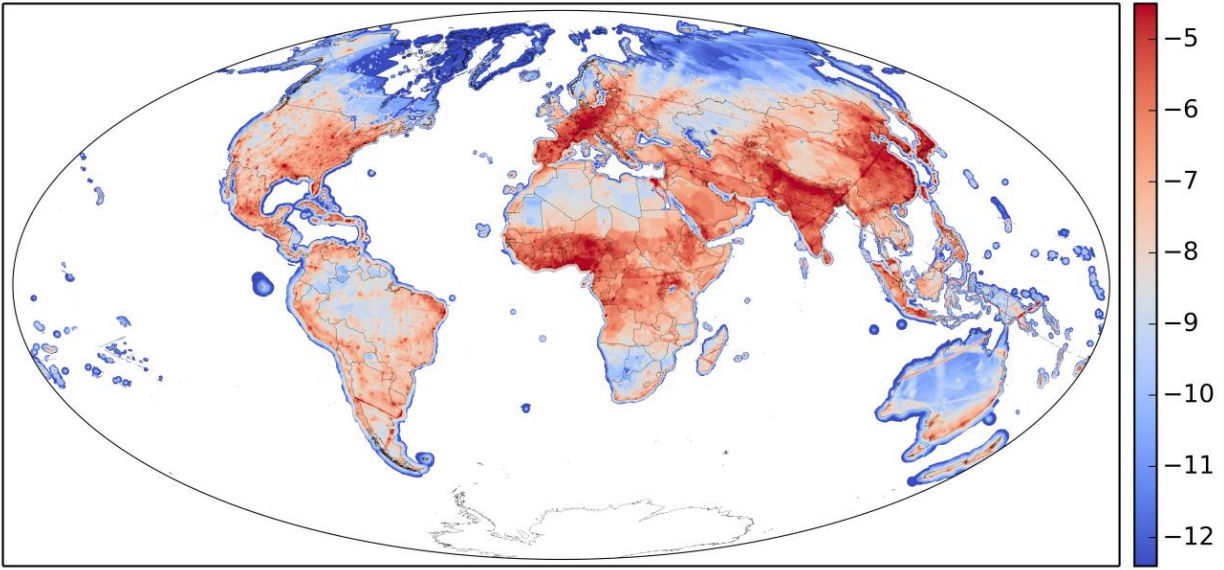

*Figure 3: Advanced asteroid risk map. This combines impact probability, exposed population and the vulnerability of that population based on a physical impact effect calculation. The colour in each region indicates the risk level. Risk is normalized with respect to global risk and is colour coded using a logarithmic scale.*

To assess the significance of physical impact effects on the risk distribution, the global asteroid risk distribution was recalculated considering impact effects and the advanced risk distribution is shown in Figure 3. Noticeable differences between Figure 2 and Figure 3 are that some individual asteroid corridors disappear while others become more pronounced because they correspond to higher energy impacts (larger asteroids) that raise the risk in the areas that are crossed by these corridors. Furthermore, the risk landscape extends beyond continental territory and beyond the shores of islands because physical effects of near coastal impacts propagate onto closeby land (aerodynamic shockwaves and thermal radiation).

Finally, the national risk list based on risk results with and without physical impact effects was produced for quantitative analysis of changes in global risk distribution. Figure 4 shows the risk list and also includes the relative global population share that each nation accommodates. The data are sorted by population size of the 40 most populous nations in descending order and population size is a good proxy for risk. In fact, national population size and risk values share a correlation coefficient of 0.953 for the simplified results and 0.907 for the advanced results.

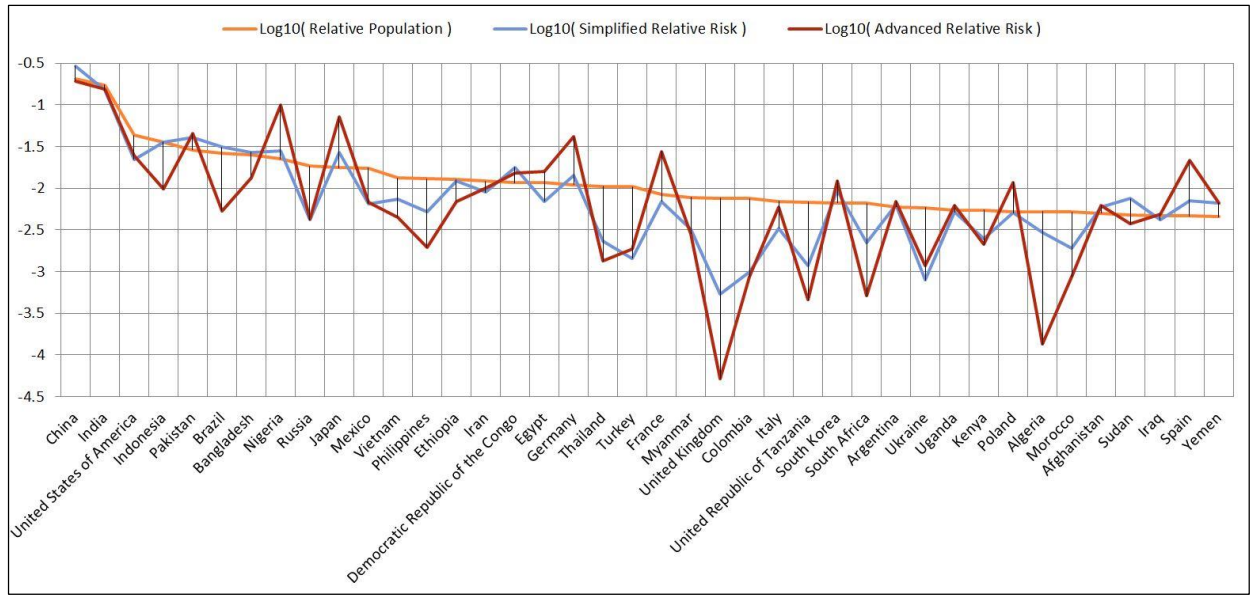

*Figure 4: National risk list for the 40 most populous countries. Risk estimates are provided both with and without taking account of physical impact effects. Both sets of estimates focus on direct land impacts (and near coastal impacts in the case of the new results). The risk estimates are presented on a logarithmic scale and based on relative values that express the global share of risk and population accredited to each nation.*

## 4. DISCUSSION

The first, general observation, based on the risk list (Figure 4), is that population size is a close proxy for national direct asteroid impact risk. The risk data share large correlation coefficients with population data of 0.953 and 0.907 in the cases without and with impact effects, respectively. The inclusion of physical impact effects did not have a major impact on this rule of thumb. This assessment is further supported by visual inspection of the global risk distribution maps of Figure 2 and Figure 3. While the appearance of individual impact corridors varies, the occurrence of broader, high risk areas remains constant such as central Africa, India, China's coastal region and central Europe.

The close correlation of population size with direct impact risk distribution stems from the generally recognised assumption that over timescales measured in hundreds of years, the impact distribution of asteroids on Earth is approximately uniform [19]. However, two aspects result in a non-uniform risk distribution in the simplified case: First, the asteroid sample size is limited and results in a non-uniform impact distribution. Second, and more importantly, impact probabilities differ between the impact instances of the asteroids and, thus, their impact corridors carry varying weights in the risk calculation resulting in additional departure from uniformity [19]. If all asteroids had the same impact probability and if the sample size were infinite, the impacts in this analysis would also be uniformly distributed. Furthermore, if impact effects are disregarded, as has been done in the simplified results, the risk distribution would mirror the population distribution perfectly in the long-term (i.e., under the assumption of a uniform impact distribution). Consequently, the variation between population and risk estimates in the simplified analysis is a function of individual impact probabilities of the VIs and, to a lesser extent, limited sample size. These two factors account for the deviation in distribution that results in a correlation coefficient of 0.953 (instead of 1.0) between population and simplified risk. Importantly, the disregard for physical impact effects ignores the fact that the analysed asteroids differ in size and impact speed. Instead, the consequences of all impacts were treated equally by assuming that only the population in the impact map cell is counted as casualties.

With the inclusion of physical impact effects the above assumption of equal impact consequences (effectively disregarding physical properties of each asteroid such as size and speed) is corrected. On one hand, a national risk increase in the advanced analysis reflects that this nation happens to be affected by a high energy impact, while, on the other hand, a decrease means that this nation is affected only by low energy impacts with fewer consequences.

Through the inclusion of physical impact effects, additional complexity is added to the analysis because the analysed asteroids differ in size as well as in impact speed and risk varies accordingly. Consequently, the advanced risk estimates reflect this additional complexity through a larger difference between population density and the advanced risk result. Noticeably, the correlation coefficient decreases from 0.953 to 0.907. This finding is confirmed in the risk list of Figure 4 where risk numbers that account for impact effects deviate more from population than those disregarding physical impact effects.

The increase in variation due to the inclusion of impact effects is moderate as expressed by the decreasing correlation coefficient (only 4.8% with respect to previous results) and might be weaker than expected. This outcome can be partially ascribed to the lack of tsunami modelling as it is expected that nations with a disproportionately long coastline relative to their country size would accumulate higher risk values if tsunamis were included [11].

Figure 4 is ordered by population size. In reference [25], a similar list is sorted by decreasing risk presenting the 40 countries that experience highest risk based on results that do not account for impact effects. The inclusion of impact effects caused a substantial reshuffling in the risk ranking. Brazil, for instance, dropped by 32 positions from 5 to 37 (in a list of 206 countries). On the other hand, Russia climbed by 9 positions from 37 to 28. The median change in position in the 40 most populated countries was 9.5 ranks. These findings show that, to predict the specific risk faced by a country, models need to account for impact effects because predictions that are based solely on impact probability and population produce only moderately accurate risk estimates. While the assertion that population size serves as a proxy for asteroid risk remains true when considering the near-equally distributed background risk of asteroid impacts, analysis that includes physical impact effects as well as the spatial distribution of impact locations is needed to estimate risk levels for individual countries.

The asteroid lists maintained by ESA and NASA change over time with the discovery of new probability asteroids and the exclusion of asteroids that are certain to miss the Earth based on new observations. New observations will also adjust the impact probability of asteroids already present in the lists. Only about 1% of all Near Earth Asteroids have been observed [26] and the majority, especially in the sub-km size regime, have yet to be discovered. Consequently, the results presented here represent only a snapshot in time of the known asteroid hazard and the risk landscape will change over time. However, the conclusions drawn based on the current analysis are independent of the characteristics of individual asteroids and the conclusions are, thus, expected to hold true in the future.

The results show that population size is an indicator for impact risk and, thus, existing tools that can calculate impact locations may be used to obtain an intuitive understanding of risk distribution. However, to assess the risk of individual countries or the risk posed by individual asteroids, asteroid impact effect modelling is a necessity.

## 5. CONCLUSIONS

This paper addresses the significance of impact effect modelling when estimating the global asteroid risk distribution. The impactor sample used was made up of the impact corridors and probability distributions of 261 VIs belonging to 69 asteroids that currently have a chance of colliding with the Earth before 2100. The distributions were calculated and visualized using the ARMOR software tool that can project impact probabilities of known asteroids onto the surface of the Earth and estimate impact effects as well as risk.

The predicted risk landscape changed moderately with the inclusion of physical impact effects. Furthermore, the observation from previous results that population size is a suitable proxy for national risk remains valid after the introduction of impact effects. The relationship between population size and relative risk reduced by 4.8% when physical impact effects were introduced, and the population-risk correlation coefficient reduced from 0.953 to 0.907. The reason for the decrease in correlation is that physical impact effects add complexity to the analysis yielding greater variation in the risk estimates. As a rule of thumb, population size helps to estimate national risk, but high impact probability asteroids or large asteroids can significantly alter this expectation. Therefore, when considering the abstract background threat of asteroids, population size serves as a suitable proxy to identify those countries that would suffer most casualties. Additionally, it might be possible to obtain a qualitative understanding of risk distribution along the possible impact locations of an asteroid using existing tools that can calculate impact locations. However, when facing a concrete threat, population size is insufficient to identify which country is most at risk and more detailed analysis including physical impact effect modelling is needed.

ACKNOWLEDGEMENTS

The work is supported by the Marie Curie Initial Training Network Stardust, FP7-PEOPLE-2012-ITN, Grant Agreement 317185. Peter Atkinson is grateful to the University of Utrecht for supporting him with The Belle van Zuylen Chair.

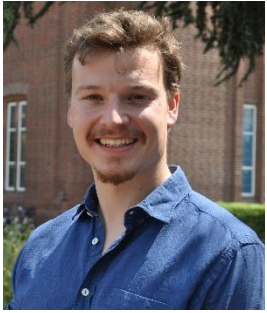

In 2012, **Clemens Rumpf** received his "Diplom-Ingenieur" aerospace engineering degree from Braunschweig University of Technology in Germany and Purdue University in the US. He has work experience in GNC (Kalman filtering) at the DLR Institute of Space Systems in Bremen. Between 2012 and 2013, he was part of the ESA Lunar Lander team at ESTEC as a GNC and space system engineer. He is currently a Research Fellow of the European Stardust network and pursuing a PhD in asteroid risk assessment at the University of Southampton in the UK where he initiated and leads a project to build the University's first CubeSat.

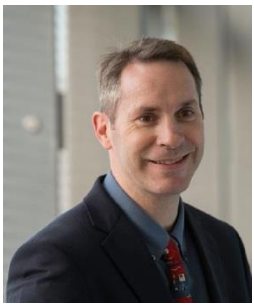

**Hugh G. Lewis** gained a Masters degree in Control Systems from the University of Sheffield in 1993 and a PhD in Remote Sensing from the University of Southampton in 1998.

Today, he is a Senior Lecturer in Aerospace Engineering at the University of Southampton and leads the Astronautics Research Group's space debris activities. He is a member of the UK Space Agency delegation to the Inter-Agency Space Debris Coordination Committee (IADC) and the Space Missions Planning Advisory Group, and he is also a member of the UK delegation to the United Nations Committee on the Peaceful Uses of Outer Space.

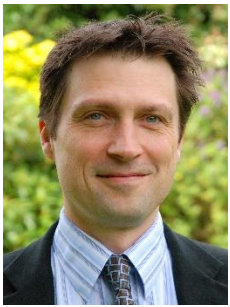

**Peter M. Atkinson** received the BSc degree in Geography from the University of Nottingham in 1986 and the PhD degree from the University of Sheffield (NERC CASE award with Rothamsted Experimental Station) in 1990. More recently, he received the MBA degree from the University of Southampton in 2012. He is currently Dean of the Faculty of Science and Technology at Lancaster University. From 2002 to 2015 he was full Professor of Geography at the University Southampton. His main interests are in the development and application of remote sensing and spatial statistical techniques to a wide range of environmental and epidemiological hazards. Prof. Atkinson has published around 200 peer-reviewed articles in international scientific journals.